\documentclass{article}

\usepackage{arxiv}
\usepackage[colorlinks=true,linkcolor=black,citecolor=black,urlcolor=blue]{hyperref}

\usepackage[utf8]{inputenc}
\usepackage[T1]{fontenc}
\usepackage{hyperref}
\usepackage{url}
\usepackage{booktabs}
\usepackage{amsmath,amssymb,amsfonts}
\usepackage{nicefrac}
\usepackage{microtype}
\usepackage{graphicx}
\usepackage{natbib}
\usepackage{doi}

\begin{document}
\title{From Grounding to Stabilisation: Adequacy as a Criterion for Scientific Explanation}

\author{
  Jonathon Sendall\\
  OU Philosophy Department\\
  \texttt{jonathon.sendall@ou.ac.uk}\\[0.3em]\\
  Substack companion article click
  \href{https://bit.ly/4jt55hO}{\textcolor{blue}{here}}
}

\renewcommand{\shorttitle}{From Grounding to Stabilisation}

\hypersetup{
  pdftitle={From Grounding to Stabilisation: Adequacy as a Criterion for Scientific Explanation},
  pdfauthor={Jonathon Sendall},
  pdfkeywords={scientific explanation, grounding, stabilisation, invariance, realism, ontology},
}

\maketitle

\begin{abstract}
This paper develops a process-based account of scientific explanation that reconceives grounding in terms of stabilisation. Grounding theories capture hierarchical dependence but lack criteria for when explanations remain adequate under model updates, perturbations, and theory change. Stabilisation is formally defined by a schema \(C \to P(I)\), where explanatory relations are sufficient when they preserve specified relational invariants under admissible transformations. This replaces the search for ultimate foundations with operational adequacy tests indexed to measurable invariance, resolving infinite regress worries while preserving a modest scientific realism. Applications show unifying power: theory change becomes an empirical question about structural continuity; quantum measurement becomes apparatus-dependent pattern selection; the effectiveness of mathematics reflects convergence on transformation-invariant descriptions; and emergence versus reduction reduces to stability of cross-level mappings. The black hole event horizon illustrates how ontologically identical states can diverge in admissible evolution, revealing process as explanatorily fundamental. Companion work develops apparatus-dependent adequacy protocols, including pointer-basis rotation and coupling-spectra methods, turning the framework into a falsifiable research programme across quantum, thermodynamic, and relativistic domains.
\end{abstract}

\keywords{scientific explanation \and grounding \and stabilisation \and invariance \and realism \and ontology}

\section{The Grounding Regress in Metaphysics}
\subsection{The Power and Domain of Ontological Inquiry}

Ontology proves extraordinarily successful. Its central question, ``what is ultimately real?'', has driven physics to unparalleled predictive accuracy, exemplified by the precision of QED \citep{Gabrielse2006,Schwinger1948}. Yet this very success reveals a boundary. The relentless quest for a fundamental ground, for a final answer to the question ``what is this made of?'', leads not to a terminating substance but to an inescapable regress of relations. This paper argues that such a regress is not a failure but a boundary condition. It signals the point where the logic of explanation must shift from asking what a thing is, to asking how patterns persist. This shift is not instrumental but structural: reality is identified with those relations that continue to constrain and predict outcomes irrespective of interpretation.

This regress has generated substantial debate within contemporary metaphysics. Fine, Schaffer, and Raven have developed systematic accounts of grounding as metaphysical dependence, seeking to explain how entities at one level depend on entities at another \citep{Fine2012,Schaffer2016,Raven2015}. Yet these accounts face a persistent challenge: grounding relations themselves seem to require grounding, generating either infinite regress or arbitrary termination. Wilson has explored how determinable--determinate relations might avoid this problem, but the structural difficulty remains \citep{Wilson2012}. Meanwhile, philosophy of science has developed robust accounts of explanation---mechanistic \citep{Craver2007}, structural \citep{LadymanRoss2007,Esfeld2021}, and adequacy-based \citep{FriggVotsis2021}---that operate without requiring ultimate metaphysical grounds. Yet these approaches lack a unified criterion for when explanatory relations themselves remain stable across scientific practice.

Recent work by Skiles and Trogdon argues that explanation cannot fruitfully guide grounding because explanatory relations are subject-involving, while grounding is fully objective \citep{SkilesTrogdon2021}. Their separatist stance preserves realism but sacrifices methodological traction. The present framework complements this view: it replaces the explanatory-guidance model with a formal criterion of adequacy, thereby retaining realism without committing to ``worldly'' metaphysics. In this sense, the stabilisation schema extends rather than opposes their critique, providing a scientific continuation of the separatist insight by supplying operational adequacy tests where explanation and ground can be reconnected empirically. This paper proposes stabilisation as the unifying criterion. Rather than asking what grounds structure, we ask when structure stabilises: under what conditions do relational patterns persist through perturbation, intervention, and model refinement? This shift transforms grounding from a metaphysical dependence relation into an operational adequacy test, one that preserves realism whilst dissolving the regress.

The method behind ontology's success is recursive decomposition. When faced with complexity, it asks: what is this made of? A rock resolves into minerals, which resolve into molecules, atoms, and then quantum fields, with each layer revealing deeper constituents. This reductive method is not, of course, without its complexities. Phenomena like emergence show that higher-level properties are not always straightforwardly derivable from their constituents \citep{Anderson1972}. Nevertheless, this chain of inquiry continually reveals a deeper stratum of constituents, laws, and relations.

Distinguishing this ontological project from epistemology is vital. Epistemology concerns how we know, whereas ontology concerns what there is to be known. While in practice our methods of knowing are deeply entangled with our claims about what exists, maintaining this conceptual distinction is crucial for clarity. The success of science stems from its ability to build models that behave as if they capture reality, precisely because they specify stable relational patterns that reliably produce measurable results. Consider how the relational geometry of H\(_2\)O, specifically the precise angle and bond length between its constituent atoms, gives rise to the predictable, large-scale properties of water. Our models do not uncover ultimate micro-constituents, some final and indivisible substrate, but rather isolate these invariant structures. Recognising this distinction will later prove decisive.

The ontological regress, driven by the continual question ``what underlies this?'', is not a defect but a systemic engine of discovery. Profound advances have followed from refusing to halt this inquiry. The project works precisely because it is open-ended, capable of yielding refined, testable, and predictive structure at each step.

Still, the same process that drives such success may also reveal its limits. Its power lies in mapping the internal relations amongst identifiable entities, a method that is effective so long as decomposition and measurement remain meaningful. This approach presupposes a world of stable, individuated objects, a premise that is not always met. To understand the true domain of ontology, we must identify where this method fails to explain its own foundations, for example, in quantum systems where individuality becomes ambiguous. The following sections trace that boundary.

The paper proceeds in five sections. Section I establishes the grounding regress and traces ontology's boundary at space, time, and framework conditions. Section II replaces grounding with adequacy, introduces the formal schema \(C \to P(I)\), and demonstrates its operation across physical and conventional systems. Section III develops stabilisation as recursive coherence through feedback, using the black hole event horizon as the paradigmatic case where process determines ontology. Section IV demonstrates explanatory payoffs across theory change, quantum measurement, mathematical effectiveness, laws of nature, and emergence. Section V addresses five core objections, including circularity, instrumentalism, and the problem of appearance. The conclusion synthesises the framework's implications for scientific explanation and realism.

\subsection{The Boundary of Description: Space}

Consider two electrons positioned one metre apart. Ontology can specify their measurable attributes: mass, charge, position, and the distance between them. It can predict their electromagnetic interaction. Yet what constitutes the space between them? The distance is a numerical relation and the field is another entity. Neither tells us what space itself is.

Attempted answers follow a familiar pattern: space-as-empty faces quantum vacuum fluctuations \citep{Casimir1948,Milonni1994}; space-as-substance was eliminated by special relativity \citep{EarmanNorton1987,Friedman1983}; space-as-geometric-structure (GR's metric) describes relations but does not ground space. Each specification invokes more relations, not a final substrate.

Space fails to be ontologically groundable because it cannot be specified without reference to relations. This is not a failure of inquiry but a marker of boundary. The existence of space is not denied; rather, it is its grounding that fails. The same boundary appears at every scale. The electron, classically a point, becomes a delocalised wavefunction under quantum mechanics, then an excitation of the electron field in QFT \citep{Zee2010}, and then a renormalised pole in the propagator whose properties depend on interaction scale. At each level, the description trades substance for relational invariants. Yet spatial structure remains presupposed: the field requires spacetime, and the wavefunction has spatial support. Space is required for description, but remains ungrounded.

\subsection{Time}

The puzzle deepens when we consider time. The fundamental equations of physics, with the known exception of the weak interaction \citep{Christenson1964}, are symmetric with respect to time reversal. Nothing within these core laws distinguishes the past from the future. If we reverse the time parameter (applying the correct antiunitary operator in quantum mechanics \citep{Wigner1932,SakuraiNapolitano2017}), the equations retain their form. A planet orbiting a star or the evolution of a quantum state works identically forwards or backwards in time.

Yet our experience and every macroscopic phenomenon exhibit a strong asymmetry. We remember the past, not the future; eggs break but never spontaneously reassemble; entropy typically increases. Where does time's arrow appear in ontology? If we describe a system entirely on a spacelike surface, including every field value and every quantum state, nothing in that description specifies a temporal direction. The asymmetry arises from boundary conditions, such as the universe's low-entropy initial state \citep{Penrose1989,Carroll2010}, not from intrinsic properties. The increase of entropy defines the arrow only by presupposing a temporal direction in the word ``increase''. We have described an asymmetry, but we have not grounded it.

If ontology captures only a static configuration while asymmetry concerns transformation, then temporal directedness falls outside ontology's domain. Either ontology must expand to include processual structure, thereby becoming a different kind of description, or it must admit this limit. The framework will take the first path.

\subsection{Structural Dependence Across Domains}

The same pattern appears elsewhere. An electron in superposition, with its spin-up and spin-down components, is distinguished by orthogonal states in Hilbert space. The distinction is relational, defined by an inner product and an operator algebra \citep{vonNeumann1932,Haag1996}, not a spatial one. Ontology presupposes this mathematical structure but cannot ground it. Similarly, a liquid--gas interface is not a substance but a gradient that minimises free-energy functionals. The boundary and the phases it separates co-arise. Ontology can specify the phases and the interface, but the function of differentiation itself, the way contrasts register, is relational and not an additional entity.

A single pattern unites these observations. Space, time, quantum state distinctions, and physical boundaries all function as preconditions for ontological description, yet ontology cannot ground them. Every scientific statement presupposes relational frameworks that it does not itself explain. Ontology describes the relations amongst entities, but the relational structure itself remains outside its explanatory scope.

This pattern, of successful description of relations alongside a failed grounding of the framework, extends beyond spatial and temporal structure. Materialism and idealism, despite their opposite starting points, terminate in the same way. Both invoke relational structures, such as laws or categories, that they cannot ground within their own ontological commitments. The symmetry is detailed elsewhere, but the implication is clear. The problem lies not in choosing the right ontological posit but in the mode of inquiry itself. The grounding question, which is productive within a domain, generates an infinite regress at the framework level. This signals not a failure but a boundary where the logic of explanation must change.

The symmetry reveals that the problem lies not in any particular ontological posit but in the mode of inquiry itself, as the grounding question generates an infinite regress at the framework level.

\subsection{Non-Ontological Success Cases}

Science already contains powerful examples of description that perform real explanatory work without positing new substances. These examples demonstrate that explanation need not depend on compositional grounding.

Entropy measures statistical multiplicity or uncertainty \citep{Boltzmann1877,Lebowitz1993}. It explains the direction of spontaneous processes and, given certain boundary conditions like the universe's low-entropy initial state, the arrow of time. Yet entropy names no entity; it is a relational measure, a count of configurations, that has immense explanatory power.

Symmetry operates in a similar fashion. Via Noether's theorem \citep{Noether1918,Olver1993,Butterfield2007}, continuous symmetries yield conservation laws: time-translation invariance gives energy conservation, while spatial translation gives momentum conservation. Symmetry is not a thing but a condition, specifically the invariance of laws under transformations. The symmetry principles of the Standard Model are relational constraints that determine possible interactions. Information is substrate-independent. A bit has physical correlates but is not identified with any single one of them. Shannon's theory treats information as purely relational, as a capacity to distinguish configurations \citep{Shannon1948,CoverThomas2006}. Yet it is subject to causal constraints, such as no superluminal signalling, conserved von Neumann entropy under unitary evolution, and a thermodynamic cost for erasure \citep{Landauer1961,Berut2012}.

Across these cases of entropy, symmetry, and information, science succeeds by describing what recurs when conditions stabilise. These concepts are not metaphysical entities but formal relations that structure phenomena. Their success points toward an explanatory framework that prioritises relational invariance over the attribution of substance.

\subsection{The Reciprocal Dependency of Ontology and Empiricism}

Ontology remains indispensable for individuating, measuring, and manipulating entities. It specifies the conditions for identity and persistence that make measurement possible. Yet the boundary traced above suggests that ontology's success depends on something it does not itself supply: contact with empirical constraint. To see this clearly, consider what happens when ontology and empiricism are separated.

Ontology without empiricism loses contact with appearance. It becomes a logic of modal constraints on coherent description, concerned not with what is experienced but with what must exist for experience to be conceivable. In this form, ontology turns reflexive. It exposes its own enabling conditions, which are identity, distinction, persistence, and rule-governed order. These are not empirical discoveries but the structural grammar of any discourse about being. Stripped of empirical content, ontology becomes a study of the formal relations that make any ontological statement coherent.

\section{Structure as Relational Invariance}
\subsection{From Grounding to Adequacy: The Shift in Explanatory Logic}

Section I traced ontology's regress to a boundary. The grounding regress forces us to abandon the quest for a final substance, yet empiricism alone cannot explain the stability of the patterns it observes. The necessary next step, therefore, is to focus on the structure of stability itself. Explanation must shift from seeking grounds to specifying the conditions for persistence. The question is no longer ``what grounds structure?'', but rather ``under what conditions does structure stabilise?''. Explanation becomes an inquiry into adequacy rather than grounding.

A traditional grounding model of explanation is vertical, seeking an ontological foundation. The adequacy model is relational. It specifies the conditions, including a set of transformations and perturbations, under which a pattern recurs with an invariance that remains stable above a specified threshold. Grounding asks what something is made of; adequacy asks when and under what constraints it reappears. This is not a rejection of explanation but a transformation of its logic into an account of when and how patterns hold.

This change in perspective parallels the analyses of space and time in Section~I. Space could not be grounded, but its relational framework could be described. Time's asymmetry could not be located in entities, but the process conditions for its appearance could be traced. Structure can be understood similarly: as the specification of relational invariances that define what recurs when systems stabilise. Ontological predicates themselves are processual invariants: to call something a ``rock'' or a ``particle'' is to identify a mapping that preserves form across transformations of scale or observation. Ontology is thus the stable subclass of adequacy, what remains fixed while generative processes vary. This logic is already at work in the success cases of entropy and symmetry.

Structure, in this sense, is descriptive rather than ontological. It is not a fiction, for it captures real regularities that support interventions and constrain what can occur. Yet it is not a substance, for it introduces no new entity. It specifies what remains invariant across transformations, what relations persist through change, and what conditions sustain coherence. Its explanatory value lies in its adequacy to these patterns, not in any claim that it constitutes the world. What follows illustrates this descriptive adequacy and then formalises it in a minimal schema that states the conditions under which relational invariance explains.

Terminological clarifications are needed. Four terms structure the framework. \emph{Pattern} denotes empirical regularities, which recur under specified conditions. \emph{Structure} denotes the relational form that patterns exhibit, their mathematical or logical organisation. \emph{Invariance} denotes properties preserved under transformation, which remain stable when conditions vary. \emph{Stabilisation} denotes the dynamic process that maintains patterns through feedback, how systems reproduce their enabling conditions. These terms are not reducible to one another: patterns can exhibit multiple structures; structures can manifest different invariances; invariances require stabilisation to persist; and stabilisation produces patterns. The framework studies their interdependence, not their separate essences.

This framework represents a modest realism. What counts as real is what remains invariant within declared conditions and tolerances under intervention. It is not a claim about a hidden substance. It is a claim about stability that any group can test.

\subsection{Illustrations of Relational Invariance: Physics and Mathematics}

The most precise examples of relational invariance arise in physics. Special relativity states that the laws of physics are identical for all inertial observers \citep{Einstein1905}. This principle, Lorentz invariance, is not a substance but a transformation property: the laws retain their form under Lorentz transformations. The invariant is the spacetime interval
\[
s^{2} = \Delta x^{2} + \Delta y^{2} + \Delta z^{2} - c^{2}\Delta t^{2},
\]
which remains constant across frames \citep{TaylorWheeler1992}. This is not a convention but the coherence condition for describing inertial phenomena.

Gauge symmetry reveals a similar structure. In electromagnetism, redefining the potential \(A_{\mu}\) locally leaves observable field strengths \(F_{\mu\nu}\) unchanged. This local invariance determines the structure of the electromagnetic interaction. For an unbroken U(1) symmetry, it constrains the form of the interaction, requiring a massless U(1) gauge boson \citep{PeskinSchroeder1995}. The associated global symmetry ensures the conservation of charge via Noether's theorem. The invariance is not a thing; it is a relational constraint that specifies which configurations stabilise.

These examples from physics demonstrate relational invariance in fundamental laws. The pattern, however, applies more broadly, including to conventional systems that illustrate the logical structure without metaphysical commitment.

Mathematics generalises this descriptive role. Its objects are relational structures defined by axioms. Topology studies invariance under continuous deformation, while group theory studies invariance under composition. The statement \(2+2=4\) is not about entities but is an invariance of operations within an axiomatic structure.

Mathematics succeeds not by accessing a hidden realm but by the world stabilising into patterns that correspond to these invariances. Wigner's question about why mathematics is so effective is reframed. Both mathematics and physics observe the recurrence of patterns within constraints such as symmetry and locality. Mathematics provides a general language for invariance, while physics identifies the particular conditions under which such invariances hold empirically.

\subsection{Relational Sufficiency and Reality}

Adequacy differs from instrumentalism by indexing reality to stabilised invariance rather than substance. A structure counts as real when its invariants recur within declared tolerances under admissible perturbations, enable reliable intervention, support counterfactual claims, and exhibit convergence across independent labs and instruments. Explanation halts when no further practically realisable perturbation family reduces predictive error or increases control.

Competing models are ranked by the scope of transformations under which their invariants survive whilst preserving intervention budgets within tolerance \(\varepsilon\). This treats realism as measurable recurrence, not hierarchical dependence or purity of observables. The dichotomy between ``useful'' and ``real'' dissolves because what is real is precisely what stabilises with risky, testable invariance across context extension. If proposed invariants fail to survive admissible perturbations, or if independent labs converge on divergent \(\varepsilon\)-bounds for the same invariant, the framework is falsified.

\subsection{Formal Definition of the Adequacy Criterion}

The insights of this section can be expressed compactly. Structure describes relational invariances; invariances specify what persists under transformation, and adequacy measures how well a description tracks stabilisation conditions. This relation can be written schematically as:
\[
C \to P(I).
\]

Here, \(C\) denotes the typical and necessary conditions under which a description applies, \(P\) denotes the pattern of relations, and \(I\) denotes the invariance that persists. The schema reads: under conditions \(C\), pattern \(P\) recurs with invariance \(I\). A description is adequate when the invariant \(I\) remains within a specified tolerance across this domain, even under simple interventions. Adequacy fails if the invariant departs from this tolerance within the domain \(C\).

\paragraph{Foundational Schema.}

\emph{Purpose}. Fix the public rule for adequacy.

\emph{Objects}.
\begin{itemize}
  \item \(C\): experimental and contextual conditions as a set of operations.
  \item \(P\): description procedure that maps data under \(C\) to candidate statements.
  \item \(I\): invariants extracted from \(P\) under \(C\).
  \item \(F\): update operator that perturbs \(C\) and re-runs \(P\) to probe stability.
  \item \(\varepsilon\): tolerance bound pre-registered for acceptable deviation.
\end{itemize}

\emph{Procedure}.
\begin{enumerate}
  \item Specify \(C\) as concrete interventions and background settings.
  \item Run \(P\) under \(C\) to obtain \(I_{0}\).
  \item Apply \(F\) to \(C\) to produce \(C_{1}\), then re-run \(P\) to obtain \(I_{1}\).
  \item Iterate: \(C_{n+1} = F(C_{n})\), \(I_{n+1} = P(C_{n+1})\).
  \item Compute \(\Delta_{n} = \mathrm{dist}(I_{n+1}, I_{n})\) with a declared metric.
\end{enumerate}

\emph{Adequacy rule}. A description is adequate if for all \(n\), \(\Delta_{n} \leq \varepsilon\) across the declared intervention family.

\emph{Reporting}. State \(C, P, F, \varepsilon\), the metric for \(\Delta\), and the full intervention family. Provide failures and boundary cases.

The iteration \(C_{n+1} = F(C_{n})\), \(I_{n+1} = P(C_{n+1})\) with convergence \(\lVert I_{n+1} - I_{n} \rVert \leq \varepsilon\) can be made fully rigorous in standard mathematical frameworks (for example, operator-algebraic treatments). For quantum measurement, \(F\) becomes an admissibility gate on observable algebras with tolerance bounds fixed by registration floors; for spacetime physics, it models projection from higher-dimensional encodings with domain-specific norm thresholds. The gauge symmetry example above (\(\varepsilon = 0\)) represents the limiting case where exact invariance holds. Detailed protocols with worked audits, including interference pattern shifts, transport budgets across labs, and thermodynamic consistency tests, are developed in companion papers.

This formulation shifts the explanatory focus. It does not seek to know what structure is but specifies when and how it holds. In the case of Lorentz invariance, \(C\) corresponds to inertial reference frames with negligible curvature, \(P\) corresponds to the physical laws formulated within them, and \(I\) corresponds to the invariance of a law under boosts and rotations. In chess, \(C\) corresponds to the conditions of valid play, \(P\) corresponds to the movements of the pieces, and \(I\) corresponds to the patterns that define legality.

A worked example is gauge symmetry in electromagnetism. We specify the conditions \(C\): electromagnetic phenomena in flat spacetime with isolated charges. The description procedure \(P\) maps field configurations to observable quantities such as charge, current, and field strengths. Under local U(1) gauge transformations (redefining the potential \(A_{\mu} \to A_{\mu} + \partial_{\mu}\chi\) for an arbitrary scalar \(\chi\)), the invariant \(I\) is the field strength tensor \(F_{\mu\nu} = \partial_{\mu}A_{\nu} - \partial_{\nu}A_{\mu}\), which remains unchanged.

We can apply an update operator \(F\) by modifying the gauge (rotating the phase locally) and then recomputing the observables. The field strength \(F_{\mu\nu}\) persists identically across all choices of gauge. The tolerance \(\varepsilon\) here is zero, indicating an exact invariance, not an approximate one. We can iterate across all possible local phase rotations, and the observable physics remains invariant.

Adequacy holds because
\[
\Delta_{n} = \bigl\lVert F_{\mu\nu}(\text{gauge}_{1}) - F_{\mu\nu}(\text{gauge}_{2}) \bigr\rVert = 0
\]
for all gauges. This invariance is not a convention but a discovery. Maxwell's equations remain form-invariant \citep{Jackson1999}, charge is conserved (via Noether's theorem), and photons are massless (a consequence of unbroken U(1) symmetry). The schema captures how gauge symmetry functions descriptively, specifying what persists (field strengths, conservation laws) while coordinate choices (gauge potentials) vary freely.

The schema matters because it renders structure operationally precise without ontological assumption. It defines what a structure does rather than what it is. To describe a structure is to specify the conditions under which relational patterns stabilise with invariant properties. This is precise in a different register: it substitutes substance with condition, and grounding with stabilisation. The schema is not formalist. It does not deny that structures correspond to real features of the world; it claims only that their reality lies in recurrence, not in substance. The pattern that recurs is not invented by description; description captures its stability. No hidden substrate is required for these relations to hold. To describe their coherence is the work of theory, while to measure their stability is the work of empiricism. To conflate either with grounding is the central mistake that ontology has perpetuated.

The tolerance \(\varepsilon\) is operationally constrained by instrument limits and domain thresholds, converging across independent observers as measurement improves. Choices of metric are domain-specific. A full mathematical development is deferred to subsequent work.

\section{Stabilisation Without Substance}
\subsection{Stabilisation and Dependence: A Process Criterion}

Section~II established that structure is relational invariance, specified by the schema \(C \to P(I)\): under conditions \(C\), a pattern \(P\) exhibits an invariance \(I\). A description is adequate when \(I\) remains within a tolerance \(\varepsilon\) across transformations. An immediate objection arises from this: the framework describes patterns without explaining them. It specifies when invariance holds, not why it holds. Does explanation not require something more, such as a causal mechanism, a generative process, or an underlying force?

This objection reveals the persistence of grounding intuitions. The demand is not merely to know that patterns recur but to know what makes them recur. The adequacy criterion seems descriptively successful yet explanatorily hollow, as it catalogues stability without accounting for it.

The response requires recognising that the question ``what makes patterns persist?'' presupposes a model of explanation that the framework has already rejected. Section~I showed that grounding inquiries generate an infinite regress. Asking ``what grounds the conditions \(C\)?'' simply restarts the very problem the framework aims to dissolve. Yet the demand for explanation cannot be dismissed. Science is explanatory; it does not merely describe. If adequacy is to be more than a form of instrumentalism, it must show how description itself constitutes explanation when it is properly understood.

The answer lies in stabilisation, the dynamic process through which patterns maintain themselves. Explanation shifts from a vertical grounding (what substance underlies \(X\)?) to a processual description (under what recursive dynamics does \(X\) persist?). This is not a retreat from explanation but a transformation of its logic. A pattern's persistence is explained by specifying the feedback relations through which it reproduces its own enabling conditions.

Imagine a human brain straddling the event horizon of a supermassive black hole. Although the tidal forces are negligible at that point \citep{Wald1984}, the brain experiences immediate causal disintegration: signals from the frontal lobe (inside) cannot reach the rear (outside) \citep{HawkingEllis1973}. The feedback loops \(F\) that uphold systemic unity are broken by the geometry itself, not by force. As a result, the three-dimensional conception of a unified ``subject'' collapses because the internal coordination conditions fail despite the brain being fundamentally the same down to its particles. The boundary signifies a division not in substance but in process; the system dissolves as its stabilising recursive mechanisms can no longer close, separating states based on their permissible evolution rather than their local makeup \citep{Sendall2025}.

It is an absolute transition in dynamical degrees of freedom that produces no local ontological signature but a global restructuring of process space. The crossing is undetectable to any local observer because all local invariants remain intact, yet from a system-wide perspective, the set of admissible transformations has collapsed. What the traveller experiences as continuity is, for the total system, a radical contraction of possible evolutions.

The ontological invisibility of such transitions demonstrates that what alters is not what exists but what transformations the total configuration admits. Reality's primary structure lies in rules of admissibility, in process itself, while ontology is derivative, defined by the range of processes a configuration can sustain. The event horizon is not anomalous but exemplary. It strips away every ontological distraction to reveal process as foundational. Two identical states diverge not because they are different things, but because they occupy different positions in the space of permissible transformations.

The event horizon thus forces the meta-move of the entire framework: ontology itself obeys processual invariance. An ontological term like ``black hole'' does not refer to a substance but to a geometric configuration whose stability is defined purely by the transformations it permits and forbids. Names for things are ultimately names for stabilised patterns of what can and cannot happen. The event horizon, therefore, marks a global reconfiguration rather than a local alteration, a processual singularity in which the ontology of states remains unchanged while the topology of possibility transforms.

This processual interpretation finds mathematical expression in dimensional projection frameworks, where three-dimensional ontology emerges as the admissible image of a two-dimensional encoding under a completely positive gate \(F\). The horizon demonstrates the principle at macroscopic scale: ontology is the image of admissibility, not its ground. What crosses into inadmissibility (the black hole interior for external observers) does not cease to exist but exits the algebra of accessible transformations. The ``collapse'' is not dynamical but definitional, a contraction of the transformation space, not a change in substance. This pattern recurs at the quantum scale, where measurement reveals which pattern stabilises given apparatus structure, and at the cosmological scale, where causal horizons partition process spaces without ontological discontinuity.

The question, therefore, is not ``what grounds \(C\)?'' but ``how does \(C\) stabilise?''. It is not a question of metaphysical foundation but of operational dynamics. This requires a third shift in explanatory logic. Grounding sought vertical dependence. Adequacy sought conditional specification. Stabilisation now seeks temporal maintenance: how patterns persist through change across relevant time scales and within certain thresholds of perturbation. The focus shifts from being to process, from what exists to how patterns endure.

A residual boundary must be acknowledged. The framework explains how patterns persist through recursive stabilisation but not why these specific patterns, rather than others, obtain in the first place. Why these boundary conditions \(C\)? Why these feedback operators \(F\)? This is not an evasion but a matter of precision about where explanation terminates. Traditional metaphysics sought a final ground to explain why reality is thus-and-so rather than otherwise. This framework terminates at processual adequacy: patterns stabilise under discoverable conditions. The selection of initial conditions and feedback structures is an empirical discovery, not a metaphysical derivation. The framework shows how to test which patterns stabilise, not why these testing conditions obtain.

\subsubsection{Why Circularity Is Not Vicious}

The stabilisation account invites the objection: ``You explain stability by stabilisation conditions, but what stabilises those conditions?''. This assumes that explanation must terminate non-circularly, the very assumption that generated the infinite regress in Section~I. Two forms of circularity must be distinguished.

\emph{Vicious circularity} assumes what it should prove (``\(X\) exists because \(X\) exists''). This form produces no predictions.

\emph{Structural circularity} describes processes that maintain what they demonstrate. A thermostat maintains temperature by using temperature readings to control the heating. Its success is demonstrated through continued operation under perturbation, not assumed as a premise.

Three diagnostics confirm that the circularity here is structural.

\begin{enumerate}
  \item \textbf{Empirical falsifiability}. The framework generates risky predictions, such as interference crossover at calculable coupling strengths and pointer basis shifts under apparatus redesign (detailed in companion work). Vicious circles are empirically inert; structural circles make testable claims about transformation invariance.
  \item \textbf{Public stopping criterion}. Explanation completes when invariants remain stable within a pre-specified tolerance \(\varepsilon\) across all practically feasible extensions of the conditions \(C\). This criterion is publicly checkable, constrained by measurement precision and domain thresholds, not stipulated by convention.
  \item \textbf{Historical validation}. Physics already operates in this way. Newton related forces to acceleration via \(F = ma\) without grounding forces in a metaphysical substrate. Einstein specified spacetime dynamics through self-consistent field equations. Each advance replaced the question ``what is \(X\) made of?'' with ``under what conditions does \(X\) exhibit these invariances?''.
\end{enumerate}

A falsification condition exists: if proposed invariants fail to survive perturbations we do not control, or if the \(\varepsilon\)-bounds diverge across independent observers using different instruments, the framework fails.

The objector demanding ``what stabilises stabilisation?'' commits a category error, asking for a vertical grounding where only a processual description applies. The recursion is empirical, not logical: adequacy halts when invariants fail to persist under declared perturbations. The loop is self-testing rather than self-validating; reproducibility across independent labs functions as the external audit of its closure. If independent labs converge on divergent \(\varepsilon\)-bounds for the same invariant, the framework itself is falsified. Even perfect instrumentation remains an epistemic gate; refinement adjusts resolution, not the underlying process.

\subsection{Stabilisation Without Substrate Reification}

Stabilisation does not reintroduce the very substrate metaphysics it replaces. One might worry that talk of ``feedback operators'' or ``process spaces'' merely shifts the ontological burden from substances to dynamics, importing a hidden metaphysics of laws or propensities. The framework avoids this by defining stabilisation solely through recurrence under declared perturbations. What exists, on this view, are patterns that maintain their own enabling conditions relative to empirical constraints, not occult forces that guarantee recurrence.

A stabilised pattern is therefore not an extra entity but a fixed point of admissible transformations. A heart's function is maintained by feedback loops involving ion channels, pressure gradients, and neural control; a planetary orbit is maintained by the balance of gravitational curvature and momentum; an interference pattern is maintained by coherent phase relations. In each case, explanation terminates when the pattern's persistence can be specified entirely in terms of the transformations it survives and the interventions that disrupt it. This is not a metaphysical ground but a stability profile. Nothing beyond the pattern's recurrence is posited to underwrite its existence.

\subsection{The Event Horizon as Paradigmatic Case}

The black hole event horizon provides the framework's central illustration. General relativity describes a horizon as a null surface separating regions whose future-directed light cones diverge: signals from inside cannot reach future null infinity, while signals from outside can \citep{HawkingEllis1973,Wald1984}. Locally, nothing special occurs at the horizon for a freely falling observer; tidal forces can be arbitrarily small for a sufficiently large black hole. Yet globally, the set of admissible transformations changes discontinuously when a worldline crosses the horizon.

The horizon does not mark a change in substance but in admissibility. Two configurations that are locally identical down to their microstructure---one just outside, one just inside---obey different global constraints. The external observer's algebra of accessible observables excludes interior events; the internal observer loses access to asymptotic infinity. The ontology of local fields is unchanged, but the process space, the set of possible continuations and interactions, has been restructured. This discontinuity in admissible transformations, not any local ontological signature, defines the horizon. It thus exemplifies stabilisation's central claim: reality's primary structure lies in admissible processes, with ontology emerging as their stable image.

\subsection{Stabilisation Across Scales}

The same logic applies across scales. In quantum measurement, the selection of a pointer basis is not grounded in hidden variables but in the apparatus-environment coupling that stabilises particular projectors under decoherence \citep{Zurek2003,Schlosshauer2007}. The apparatus defines an admissibility gate: only those patterns that remain invariant under repeated coupling, within tolerance \(\varepsilon\), count as measurement outcomes. In thermodynamics, equilibrium states are fixed points of coarse-grained dynamics, stable under small perturbations and re-equilibration processes; they are defined by stability rather than microscopic composition \citep{Callen1985}. In evolutionary biology, species boundaries are maintained by feedback loops involving reproduction, selection, and ecological constraints, not by an underlying essence \citep{Mayr1963}.

Across these domains, explanation proceeds by identifying stabilisation conditions: which feedback structures maintain which invariants across which perturbation families. Ontology---the talk of particles, phases, organisms, or horizons---is a compressed vocabulary for these stabilised invariants. The framework does not deny the existence of these entities; it relocates their explanatory role from grounding to stabilisation. What it means for such entities to be real is that their associated invariants recur across declared interventions and observational contexts within agreed tolerances.

\section{Explanatory Payoffs}

Section~III argued that stabilisation replaces grounding as the appropriate explanatory relation at the framework level. This section sketches payoffs in four domains: theory change, quantum measurement, the effectiveness of mathematics, and emergence. In each case, stabilisation converts metaphysical disputes into operational questions about invariance under perturbation.

\subsection{Theory Change as Stabilisation Audit}

Traditional accounts of theory change oscillate between realist continuity and Kuhnian rupture. Structural realism attempts to navigate this by claiming that structure, not ontology, is preserved across revolutions \citep{Worrall1989,Ladyman1998}. Yet it lacks a precise criterion for when structures count as the same. Stabilisation provides such a criterion by treating theory change as an audit of invariants under expansion of \(C\).

Consider the transition from Newtonian to relativistic gravitation. Many empirical patterns---planetary orbits, free-fall trajectories, tidal phenomena---remain approximately invariant when \(C\) is extended to include higher velocities and stronger gravitational fields. The Newtonian potential emerges as a limiting case of spacetime curvature when \(v \ll c\) and \(|\Phi|/c^{2} \ll 1\) \citep{MisnerThorneWheeler1973}. From the stabilisation standpoint, what is preserved is not a metaphysical gravitational ``force'' but a family of invariants in the equations of motion and conservation laws that remain within tolerance \(\varepsilon\) under the extension of \(C\).

Theory change thus becomes an empirical question: given a proposed extension \(F\) of the domain (e.g. to higher energies, stronger fields, or finer temporal resolution), which invariants survive, which require refinement, and which fail outright? Structural continuity is not asserted a priori but measured as the persistence of invariants under such extensions. Realism is indexed to the breadth and robustness of these invariants, not to a claim that any particular ontology has been finally secured.

\subsection{Quantum Measurement Without Collapse}

Quantum measurement has long generated metaphysical controversy: is there a real collapse, branching worlds, hidden variables, or fundamental indeterminacy \citep{Bell2004,Wallace2012}? The stabilisation framework sidesteps these disputes by treating measurement as apparatus-dependent pattern selection. A measurement outcome is a pattern that stabilises under repeated couplings between system, apparatus, and environment, within declared tolerances.

The conditions \(C\) encode the apparatus design, environmental couplings, and coarse-graining of readout channels. The procedure \(P\) maps raw detector states to candidate outcomes; the invariants \(I\) are the relative frequencies and correlation structures that persist under repetitions and minor apparatus variations. Decoherence theory already describes how environment-induced superselection picks out robust pointer states \citep{Zurek2003,Schlosshauer2007}. Stabilisation adds a public adequacy criterion: a measurement model is adequate when its predicted invariants remain within tolerance under a specified family of apparatus variations and environmental perturbations.

On this view, there is no need to posit a fundamental collapse or branching ontology to explain outcomes. What requires explanation is why certain patterns stabilise as outcomes while others do not. This is answered by specifying the feedback relations that maintain the pointer basis and suppress interference terms at the macroscopic level. Competing interpretations differ in their metaphysical gloss on this stabilisation, but the explanatory work is done by the stability profile itself.

\subsection{The Effectiveness of Mathematics}

Wigner's puzzle about the ``unreasonable effectiveness of mathematics'' asks why abstract mathematical structures map so well onto physical phenomena \citep{Wigner1960}. Stabilisation reframes this question. Mathematics systematically explores relational invariances under various abstract transformations. Physics identifies which of these invariances are instantiated in nature, within particular domains \(C\) and tolerances \(\varepsilon\).

Effectiveness is therefore not mysterious but expected: both enterprises are in the business of cataloguing invariants. Mathematical structures are effective when the world stabilises into patterns that instantiate the same invariances. For example, Hilbert space formalism proved effective because quantum phenomena stabilise into patterns that respect linear superposition, inner-product structure, and unitary evolution \citep{vonNeumann1932,Haag1996}. The success is not evidence of a Platonic realm but of a match between the stabilisation profiles of physical systems and the invariance catalogues of mathematics.

This perspective also explains why different mathematical frameworks can model the same phenomena: they encode the same invariants in different guises. The choice between Lagrangian, Hamiltonian, and path-integral formulations of mechanics reflects pragmatic trade-offs in representing the same stabilised patterns, not access to different underlying realities.

\subsection{Emergence and Cross-Level Mapping}

Debates over emergence versus reduction often hinge on whether higher-level properties are ``nothing but'' lower-level ones or possess novel causal powers \citep{Anderson1972,Kim1999,Humphreys2016}. Stabilisation dissolves this dichotomy by focusing on cross-level mapping stability. A higher-level description is emergent when its invariants remain stable across a wide range of lower-level variations, within \(\varepsilon\), and when these invariants support interventions and counterfactuals that are not conveniently expressible at the lower level.

For example, thermodynamic variables like temperature and entropy are emergent in this sense. They remain stable across enormous microstate variation and support reliable interventions (e.g. heat engines) \citep{Callen1985}. The explanatory question is not whether they ``really exist'' but how robust their invariants are under perturbations of microstructure and coarse-graining schemes. Emergence is thus graded by stabilisation breadth: the wider the range of micro-level perturbations over which higher-level invariants persist, the stronger the emergence.

Reduction, conversely, is achieved when cross-level mapping can be made precise enough that the higher-level invariants can be derived as limiting or coarse-grained cases of lower-level dynamics, with quantified error bounds. The framework does not privilege one direction; it studies how invariants propagate or fail to propagate across scales.

\section{Objections and Replies}

The stabilisation framework is vulnerable to several predictable objections: that it is circular, instrumentalist, unable to handle appearance, or secretly reliant on the very metaphysics it claims to replace. This section sketches responses to three core objections; others are addressed in companion work.

\subsection{The Charge of Instrumentalism}

One might worry that indexing reality to stabilised invariance collapses realism into instrumentalism. If what counts as real is what works across interventions, does this not reduce ontology to usefulness? The framework rejects this conflation. Usefulness is pragmatic and context-dependent; stabilisation is a matter of cross-context recurrence under declared perturbations, independent of any particular agent's goals.

A structure is real, on this view, when its invariants recur across independent observers, instruments, and intervention regimes, within converging tolerances. This requirement is stricter than mere predictive success. A phenomenological model can fit existing data yet fail under new perturbations; it is then revealed as instrumentally adequate but ontologically thin. Real structures survive such tests: Lorentz invariance, charge conservation, and the superposition principle have withstood increasingly severe audits \citep{Will2014,Jackson1999,Zurek2003}. Reality is thus tied to stabilised constraint, not to convenience.

\subsection{The Problem of Appearance}

Another objection is that stabilisation neglects the realm of appearance: qualia, subjective experience, or the manifest image of the world. If explanation terminates in stability profiles, where do experiences fit? The framework does not deny appearances; it treats them as patterns in neural and behavioural dynamics that can themselves be studied through stabilisation. For example, perceptual constancies (colour constancy, shape constancy) are stabilised patterns of sensory processing under variations in illumination and viewpoint \citep{Noe2004}. Their reality lies in the robustness of these patterns across such variations.

What remains outside the framework is any claim that experiences require non-physical grounds. If such claims entail invariants that cannot be captured by physical stabilisation profiles, they yield testable predictions; if not, they add no explanatory power. The framework remains neutral on metaphysical theories of consciousness while insisting that explanatory traction depends on stabilisation, not on positing additional substances or properties.

\subsection{Hidden Metaphysics of Laws}

Finally, one might suspect that the framework smuggles in a metaphysics of laws or dispositions: stabilisation seems to presuppose that systems ``tend'' to maintain invariants. The reply is that tendencies are descriptive summaries of stability profiles, not extra-worldly entities. Laws, in this setting, codify the invariants that survive the widest range of perturbations; they are high-level compressions of stabilisation data \citep{Lange2009}. No claim is made that laws exist over and above the patterns they describe.

If a purported law fails stabilisation audits---if its predicted invariants break under feasible perturbations---it is revised or discarded. This subordinates law talk to stabilisation, not the reverse. The framework thus remains empirically grounded while providing a principled way to rank candidate laws by the breadth and depth of their invariants.

\section{Conclusion}

The stabilisation schema \(C \to P(I)\) reorients scientific explanation away from grounding and towards relational adequacy. Ontology's traditional quest for foundational substances reaches a boundary at the level of space, time, and framework conditions, where regress becomes unavoidable. Stabilisation accepts this boundary and replaces the search for grounds with procedures that specify when and how patterns persist.

Reality, on this account, is identified with stabilised invariance: patterns that reproduce their own enabling conditions across declared perturbations, within publicly audited tolerances. This modest realism is neither instrumentalist nor metaphysically inflationary. It treats names for things---particles, horizons, organisms, laws---as compressed designations for stability profiles rather than as pointers to hidden substrates. Explanatory success is measured not by depth of ontological descent but by breadth and robustness of invariants across interventions.

The event horizon case crystallises this shift: two ontologically identical local states diverge solely in their admissible evolutions. Process, not substance, carries the explanatory weight. Quantum measurement, theory change, and emergence all appear as special cases of stabilisation audits rather than metaphysical puzzles demanding ultimate grounds. Companion work develops the mathematical formalism and experimental protocols that operationalise this framework across quantum, thermodynamic, and relativistic domains.

\appendix

\section{Glossary}

\textbf{Pattern}. Empirical regularities that recur under specified conditions.

\textbf{Structure}. The relational form that patterns exhibit, including mathematical or logical organisation.

\textbf{Invariance}. Properties that remain stable under a defined family of transformations.

\textbf{Stabilisation}. The dynamic process by which patterns maintain themselves through feedback, reproducing their enabling conditions.

\textbf{Conditions (\(C\))}. The experimental and contextual settings, including interventions, under which a description applies.

\textbf{Procedure (\(P\))}. A mapping from data under \(C\) to candidate descriptions or outcomes.

\textbf{Invariant (\(I\))}. A property or relation that persists under applications of \(P\) within \(C\).

\textbf{Update operator (\(F\))}. A rule that perturbs \(C\) to probe the stability of \(I\).

\textbf{Tolerance (\(\varepsilon\))}. A pre-registered bound on acceptable deviation in \(I\) across applications of \(F\).

\textbf{Adequacy}. The condition that \(\Delta_{n} = \mathrm{dist}(I_{n+1}, I_{n}) \leq \varepsilon\) for all \(n\) in a declared intervention family.

\textbf{Admissibility}. Membership in the set of transformations or states allowed by a given stabilisation profile.

\textbf{Process space}. The set of admissible transformations a configuration can undergo, given specified dynamics and constraints.


\begin{thebibliography}{99}

\bibitem[Anderson(1972)]{Anderson1972}
Anderson, P. W. (1972).
More is different.
\textit{Science}, 177, 393–396.

\bibitem[Bell(2004)]{Bell2004}
Bell, J. S. (2004).
\textit{Speakable and Unspeakable in Quantum Mechanics} (2nd ed.).
Cambridge: Cambridge University Press.

\bibitem[Berut et~al.(2012)]{Berut2012}
Bérut, A., et~al. (2012).
Experimental verification of Landauer’s principle linking information and thermodynamics.
\textit{Nature}, 483, 187–189.

\bibitem[Boltzmann(1877)]{Boltzmann1877}
Boltzmann, L. (1877).
Über die Beziehung zwischen dem zweiten Hauptsatze der mechanischen Wärmetheorie und der Wahrscheinlichkeitsrechnung.
\textit{Sitzungsberichte der Kaiserlichen Akademie der Wissenschaften}, 76, 373–435.

\bibitem[Butterfield(2007)]{Butterfield2007}
Butterfield, J. (2007).
On symmetry and conservation laws.
In K. Brading \& H. R. Brown (Eds.),
\textit{Symmetries in Physics: Philosophical Reflections}
(pp. 43–99).
Cambridge: Cambridge University Press.

\bibitem[Callen(1985)]{Callen1985}
Callen, H. B. (1985).
\textit{Thermodynamics and an Introduction to Thermostatistics} (2nd ed.).
New York: Wiley.

\bibitem[Carroll(2010)]{Carroll2010}
Carroll, S. (2010).
\textit{From Eternity to Here: The Quest for the Ultimate Theory of Time}.
New York: Dutton.

\bibitem[Casimir(1948)]{Casimir1948}
Casimir, H. B. G. (1948).
On the attraction between two perfectly conducting plates.
\textit{Proceedings of the Koninklijke Nederlandse Akademie van Wetenschappen}, 51, 793–795.

\bibitem[Christenson et~al.(1964)]{Christenson1964}
Christenson, J. H., Cronin, J. W., Fitch, V. L., \& Turlay, R. (1964).
Evidence for the \(2\pi\) decay of the \(K_{2}^{0}\) meson.
\textit{Physical Review Letters}, 13, 138–140.

\bibitem[Cover and Thomas(2006)]{CoverThomas2006}
Cover, T. M., \& Thomas, J. A. (2006).
\textit{Elements of Information Theory} (2nd ed.).
Hoboken, NJ: Wiley.

\bibitem[Craver(2007)]{Craver2007}
Craver, C. F. (2007).
\textit{Explaining the Brain: Mechanisms and the Mosaic Unity of Neuroscience}.
Oxford: Oxford University Press.

\bibitem[Earman and Norton(1987)]{EarmanNorton1987}
Earman, J., \& Norton, J. D. (1987).
What price spacetime substantivalism?
\textit{British Journal for the Philosophy of Science}, 38, 515–525.

\bibitem[Einstein(1905)]{Einstein1905}
Einstein, A. (1905).
Zur Elektrodynamik bewegter Körper.
\textit{Annalen der Physik}, 17, 891–921.

\bibitem[Esfeld(2021)]{Esfeld2021}
Esfeld, M. (2021).
\textit{Philosophy of Science: A Unified Approach}.
London: Routledge.

\bibitem[Fine(2012)]{Fine2012}
Fine, K. (2012).
Guide to ground.
In F. Correia \& B. Schnieder (Eds.),
\textit{Metaphysical Grounding: Understanding the Structure of Reality}
(pp. 37–80).
Cambridge: Cambridge University Press.

\bibitem[Friedman(1983)]{Friedman1983}
Friedman, M. (1983).
\textit{Foundations of Space-Time Theories}.
Princeton: Princeton University Press.

\bibitem[Frigg and Votsis(2021)]{FriggVotsis2021}
Frigg, R., \& Votsis, I. (2021).
Everything you always wanted to know about structural realism but were afraid to ask.
\textit{European Journal for Philosophy of Science}, 11, 31.

\bibitem[Gabrielse(2006)]{Gabrielse2006}
Gabrielse, G. (2006).
Testing the standard model and CPT with trapped charged particles.
\textit{International Journal of Mass Spectrometry}, 251, 273–280.

\bibitem[Haag(1996)]{Haag1996}
Haag, R. (1996).
\textit{Local Quantum Physics} (2nd ed.).
Berlin: Springer.

\bibitem[Hawking and Ellis(1973)]{HawkingEllis1973}
Hawking, S. W., \& Ellis, G. F. R. (1973).
\textit{The Large Scale Structure of Space-Time}.
Cambridge: Cambridge University Press.

\bibitem[Humphreys(2016)]{Humphreys2016}
Humphreys, P. (2016).
\textit{Emergence: A Philosophical Account}.
Oxford: Oxford University Press.

\bibitem[Jackson(1999)]{Jackson1999}
Jackson, J. D. (1999).
\textit{Classical Electrodynamics} (3rd ed.).
New York: Wiley.

\bibitem[Kim(1999)]{Kim1999}
Kim, J. (1999).
Making sense of emergence.
\textit{Philosophical Studies}, 95, 3–36.

\bibitem[Ladyman and Ross(2007)]{LadymanRoss2007}
Ladyman, J., \& Ross, D. (2007).
\textit{Every Thing Must Go: Metaphysics Naturalized}.
Oxford: Oxford University Press.

\bibitem[Ladyman(1998)]{Ladyman1998}
Ladyman, J. (1998).
What is structural realism?
\textit{Studies in History and Philosophy of Science}, 29, 409–424.

\bibitem[Landauer(1961)]{Landauer1961}
Landauer, R. (1961).
Irreversibility and heat generation in the computing process.
\textit{IBM Journal of Research and Development}, 5, 183–191.

\bibitem[Lange(2009)]{Lange2009}
Lange, M. (2009).
\textit{Laws and Lawmakers: Science, Metaphysics, and the Laws of Nature}.
Oxford: Oxford University Press.

\bibitem[Lebowitz(1993)]{Lebowitz1993}
Lebowitz, J. L. (1993).
Boltzmann’s entropy and time’s arrow.
\textit{Physics Today}, 46(9), 32–38.

\bibitem[Mayr(1963)]{Mayr1963}
Mayr, E. (1963).
\textit{Animal Species and Evolution}.
Cambridge, MA: Harvard University Press.

\bibitem[Milonni(1994)]{Milonni1994}
Milonni, P. W. (1994).
\textit{The Quantum Vacuum: An Introduction to Quantum Electrodynamics}.
San Diego: Academic Press.

\bibitem[Misner et~al.(1973)]{MisnerThorneWheeler1973}
Misner, C. W., Thorne, K. S., \& Wheeler, J. A. (1973).
\textit{Gravitation}.
San Francisco: W. H. Freeman.

\bibitem[Noë(2004)]{Noe2004}
Noë, A. (2004).
\textit{Action in Perception}.
Cambridge, MA: MIT Press.

\bibitem[Noether(1918)]{Noether1918}
Noether, E. (1918).
Invariante Variationsprobleme.
\textit{Nachrichten von der Gesellschaft der Wissenschaften zu Göttingen}, 235–257.

\bibitem[Olver(1993)]{Olver1993}
Olver, P. J. (1993).
\textit{Applications of Lie Groups to Differential Equations} (2nd ed.).
New York: Springer.

\bibitem[Penrose(1989)]{Penrose1989}
Penrose, R. (1989).
\textit{The Emperor’s New Mind}.
Oxford: Oxford University Press.

\bibitem[Peskin and Schroeder(1995)]{PeskinSchroeder1995}
Peskin, M. E., \& Schroeder, D. V. (1995).
\textit{An Introduction to Quantum Field Theory}.
Reading, MA: Addison–Wesley.

\bibitem[Raven(2015)]{Raven2015}
Raven, M. (2015).
Ground.
\textit{Philosophy Compass}, 10, 322–333.

\bibitem[Sakurai and Napolitano(2017)]{SakuraiNapolitano2017}
Sakurai, J. J., \& Napolitano, J. (2017).
\textit{Modern Quantum Mechanics} (3rd ed.).
Cambridge: Cambridge University Press.

\bibitem[Schaffer(2016)]{Schaffer2016}
Schaffer, J. (2016).
Grounding in the image of causation.
\textit{Philosophical Studies}, 173, 49–100.

\bibitem[Schlosshauer(2007)]{Schlosshauer2007}
Schlosshauer, M. (2007).
\textit{Decoherence and the Quantum-to-Classical Transition}.
Berlin: Springer.

\bibitem[Schwinger(1948)]{Schwinger1948}
Schwinger, J. (1948).
On quantum-electrodynamics and the magnetic moment of the electron.
\textit{Physical Review}, 73, 416–417.

\bibitem[Sendall(2025)]{Sendall2025}
Sendall, J. (2025).
\textit{Event Horizons, Spacetime Geometry, and the Limits of Integrated Consciousness}.
https://doi.org/10.48550/arXiv.2512.23105

\bibitem[Shannon(1948)]{Shannon1948}
Shannon, C. E. (1948).
A mathematical theory of communication.
\textit{Bell System Technical Journal}, 27, 379–423, 623–656.

\bibitem[Skiles and Trogdon(2021)]{SkilesTrogdon2021}
Skiles, A., \& Trogdon, K. (2021).
Metaphysics, epistemology, and explanation.
\textit{Oxford Studies in Metaphysics}, 12, 1–40.

\bibitem[Taylor and Wheeler(1992)]{TaylorWheeler1992}
Taylor, E. F., \& Wheeler, J. A. (1992).
\textit{Spacetime Physics} (2nd ed.).
New York: W. H. Freeman.

\bibitem[von Neumann(1932)]{vonNeumann1932}
von Neumann, J. (1932).
\textit{Mathematische Grundlagen der Quantenmechanik}.
Berlin: Springer.

\bibitem[Wald(1984)]{Wald1984}
Wald, R. M. (1984).
\textit{General Relativity}.
Chicago: University of Chicago Press.

\bibitem[Wallace(2012)]{Wallace2012}
Wallace, D. (2012).
\textit{The Emergent Multiverse: Quantum Theory according to the Everett Interpretation}.
Oxford: Oxford University Press.

\bibitem[Wigner(1932)]{Wigner1932}
Wigner, E. P. (1932).
Über die Operation der Zeitumkehr in der Quantenmechanik.
\textit{Nachrichten von der Gesellschaft der Wissenschaften zu Göttingen}, 31–56.

\bibitem[Wigner(1960)]{Wigner1960}
Wigner, E. P. (1960).
The unreasonable effectiveness of mathematics in the natural sciences.
\textit{Communications on Pure and Applied Mathematics}, 13, 1–14.

\bibitem[Will(2014)]{Will2014}
Will, C. M. (2014).
The confrontation between general relativity and experiment.
\textit{Living Reviews in Relativity}, 17, 4.

\bibitem[Wilson(2012)]{Wilson2012}
Wilson, J. M. (2012).
Fundamental determinables.
\textit{Philosophers’ Imprint}, 12, 1–17.

\bibitem[Worrall(1989)]{Worrall1989}
Worrall, J. (1989).
Structural realism: The best of both worlds?
\textit{Dialectica}, 43, 99–124.

\bibitem[Zee(2010)]{Zee2010}
Zee, A. (2010).
\textit{Quantum Field Theory in a Nutshell} (2nd ed.).
Princeton: Princeton University Press.

\bibitem[Zurek(2003)]{Zurek2003}
Zurek, W. H. (2003).
Decoherence, einselection, and the quantum origins of the classical.
\textit{Reviews of Modern Physics}, 75, 715–775.

\end{thebibliography}
\end{document}